\documentclass[journal]{IEEEtran}
\usepackage[cmex10]{amsmath}
\usepackage{amsthm}
\usepackage{amsmath}
\usepackage{amssymb}
\usepackage{amsfonts}
\usepackage{epic}
\usepackage{epsfig}
\usepackage{graphicx}
\usepackage[tight,footnotesize]{subfigure}
\usepackage{latexsym}
\usepackage{mathrsfs}
\usepackage{multirow}
\usepackage[all]{xy}
\usepackage{cite}

\def\va{{\bf a}}   \def\vd{{\bf d}}
\def\ve{{\bf e}}   
   
 \def\vn{{\bf n}}  
   
   \def\vx{{\bf x}} 
\def\vy{{\bf y}} \def\vz{{\bf z}}

\def\vA{{\bf A}}   
\def\vE{{\bf E}}

\def\vU{{\bf U}} \def\vV{{\bf V}}  \def\vX{{\bf X}} 
 \def\vZ{{\bf Z}}
\def\cA{{\cal A}}
\def\cG{{\cal G}}
\def\cS{{\cal S}}

\def\eps{\varepsilon}
\def\Cst{{\mathbb C}}

\date{}

\title{Measure What Should be Measured: \\ Progress and Challenges in Compressive Sensing}

\author{
Thomas Strohmer\thanks{T.\ Strohmer is with the Department of Mathematics, 
University of California at Davis, Davis CA-95616, USA.
This work was supported in part by the National Science Foundation via
grant DTRA-DMS 1042939 and by DARPA via grant N66001-11-1-4090.}}

\begin{document}
\maketitle

\begin{abstract}
Is compressive sensing overrated? Or can it live up to our expectations?
What will come after compressive sensing and sparsity? And what has
Galileo Galilei got to do with it? Compressive
sensing has taken the signal processing community by storm.
A large corpus of research devoted to the theory and numerics of
compressive sensing has been published in the last few years.
Moreover, compressive sensing has inspired and initiated intriguing new
research directions, such as matrix completion.
Potential new applications emerge at a dazzling rate. Yet some
important theoretical questions remain open, and
seemingly obvious applications keep escaping the grip of compressive
sensing. In this paper\footnote{This is not a regular IEEE-SPL paper, but
rather an invited contribution offering a vision for key advances in
emerging fields.} I discuss some of the recent progress in
compressive sensing and point out key challenges and opportunities
as the area of compressive sensing and sparse representations keeps evolving.
I also attempt to assess the long-term impact of compressive sensing.

\end{abstract}

\section{Introduction}

{\em ``Measure what can be measured''}, this quote often attributed to 
Galileo Galilei, has become a paradigm for scientific discovery that 
seems to be more dominant nowadays than ever before\footnote{The full quote says 
{\em ``Measure what can be measured and make measurable what cannot be measured''}, 
but it is disputed whether Galilei ever said or wrote these words~\cite{Kle09}. 
Nevertheless, the quote is widely accepted as a very
fitting characterization of the leitmotif of Galilei's work with respect to the
central role of the {\em experiment} in the {\em Nuova Scienza}.}.
However, in light of the data deluge we are facing today, it is perhaps time to
modify this principle to {\it ``Measure what should be measured''}.
Of course the problem is that a priori we often do not know what we should
measure and what not. What is important and what can be safely ignored? 

A typical example is a digital camera, which acquires in the order of a
million measurements each time a picture is taken, only to dump
a good portion of the data soon after the acquisition through the
application of an image compression algorithm.
In contrast, {\em compressive sensing} operates under the premise that 
signal acquisition and data compression can be carried out
simultaneously: {\em ``Measure what should be measured!''}

On the one end of the spectrum of scientific endeavour, the concept of 
compressive sensing has led to the development of new data acquisition 
devices. On the other 
end, the beauty of the underlying mathematical theory has attracted even pure 
mathematicians. And ``in between'', scientists from physics, astronomy,
engineering, biology, medical image processing, etc.\ explore the possibilities 
of sparse representations and the opportunities of compressive sensing.
It is therefore only natural for such a timely journal as
the IEEE Signal Processing Letters, that compressive sensing and sparsity
are now incorporated into the new EDICS categories.

\medskip
At the mathematical heart of compressive sensing lies the discovery that it is 
possible to reconstruct a sparse signal exactly from an underdetermined linear 
system of equations {\em and} that this can be done in a computationally
efficient manner via convex programming. To fix ideas and notation,
consider $\vA \vx=\vy$, where $\vA$ is an $m \times n$ matrix of rank $m$
with $m < n$.
Here, $\vA$ models the measurement (or sensing) process, $\vy \in \Cst^m$ is 
the vector of observations and $\vx \in \Cst^n$ is the signal of interest.
Conventional linear algebra wisdom tells us that in principle the number 
of measurements $m$ has to be at least as large as the signal length $n$, 
otherwise the system would be underdetermined and there would be infinitely 
many solutions. Most data acquisition devices of current technology obey this 
principle in one way or another (for instance, devices that follow 
Shannon's Sampling Theorem which states that the sampling rate must 
be at least twice the maximum frequency present in the signal).

Now assume that $\vx$ is sparse, i.e., $\vx$ satisfies 
$s:=\|\vx\|_0 \ll n$ (where $\|\vx\|_0 := \# \{k: x_k \neq 0\}$), 
but we do {\em not} know the locations of the non-zero entries of $\vx$. 
Due to the sparsity of $\vx$ one could try to compute
$\vx$ by solving the optimization problem
\begin{equation}
\underset{\vz}{\min} \|\vz\|_0 \quad \text{s.t.}\,\,\, \vA \vz = \vy.
\label{L0}
\end{equation}
However solving~\eqref{L0} is an NP-hard problem and thus practically not
feasible. Instead we consider its convex relaxation 
\begin{equation}
\underset{\vz}{\min} \|\vz\|_1 \quad \text{s.t.}\,\,\, \vA \vz = \vy,
\label{L1}
\end{equation}
which can be solved efficiently via linear or quadratic programming
techniques.
It is by now well-known that under certain conditions on the matrix $\vA$
and the sparsity of $\vx$, both~\eqref{L0} and~\eqref{L1} have the same
unique solution~\cite{CanRomTao,CanTao,Don}. The {\em Restricted Isometry
Property} (RIP) and the {\em coherence} of a matrix are to date the most
widely used tools to derive such conditions. Indeed, for a properly chosen
$\vA$ about $m=s \log n$ measurements suffice to uniquely recover $\vx$ 
from~\eqref{L1}. In other words, a sparse signal can be sampled at almost 
its ``information rate'' using non-adaptive linear measurements.

\medskip
Compressive sensing took the signal processing community by storm.
As the graph in~\cite{Ela12} shows, the number of publications dealing
with sparse representations and compressive sensing has grown rapidly
over the last couple of years. Admittedly, we were in a somewhat lucky 
situation when compressive sensing arrived
on the scene: Researchers in signal processing, applied harmonic analysis,
imaging sciences, and information theory had already 
fostered a culture of close collaboration and interaction over the last 
two decades or so, laying the foundation for a strong willingness from 
engineers, statisticians, and mathematicians to cooperate and learn from 
each other. This fact definitely contributed to the very quick adoption of
compressive sensing by the various research communities.

Is compressive sensing overrated?  Will compressive sensing revolutionize 
data acquisition? Can compressive sensing live up to our (admittedly, 
rather high) expectations? What are the most promising applications?
Are there still interesting open mathematical problems? 
And what will come after compressive sensing and sparse representations? 
While this article may not be able to provide satisfactory
answers to all these questions, it is nevertheless strongly motivated by 
them. I will discuss open problems and challenges, and while doing so,
shed light on some recent progress. I will also attempt to evaluate the impact 
of compressive sensing in the context of future scientific developments.

I also want to draw the reader's attention to the enlightening article 
{\em ``Sparse and Redundant Representation Modeling --- What Next?''} 
by Michael Elad in the very
same issue of this journal. I have tried to keep the topics discussed in my
article somewhat complementary to his, but, naturally, our two articles do
overlap at places, which was in part not avoidable, since we were writing
them at about the same time. The reader, who wonders why I did not mention 
the one or the other important open problem, may likely find it very eloquently 
described in Elad's paper. I want to stress at this point that the areas 
of compressive sensing and sparse representations obviously have a strong 
overlap, but one should not conflate them completely.

I assume that the reader is familiar with the basics of compressive sensing 
and sparse representations.  Excellent introductions to compressive sensing 
are the review articles~\cite{Bar07,CW08}, the soon-to-be-published 
book~\cite{FR12}, and of course the original research 
papers~\cite{CanTao,CanRomTao,Don}. A great source for sparse and redundant 
representations is~\cite{Ela10}. The reader who wants to get an overview 
of recent developments in these areas should also check out Igor Carron's 
informative blog {\em Nuit Blanche} (http://nuit-blanche.blogspot.com). 

\section{Progress and Challenges} \label{s:progresschallenges}

In this section I will discuss some problems which I consider important 
future research directions in compressive sensing. They range from very 
concrete to quite abstract/conceptual, from very theoretical to quite 
applied. In some of the problems mentioned below we already have seen 
significant progress over the last few years, others are still in their
infancy. The ordering of the problems does not reflect their importance, 
but is chosen to best aid the narrative of the paper. The list is by no
means exhaustive, moreover it is subjective and biased towards the author's
background, taste, and viewpoints. To highlight the connection with
the new EDICS related to sparsity and compressive sensing, I am listing 
the EDICS most relevant for each subsection:
Subsection~\ref{s:structured}: MLAS-SPARS, SAM-SPARCS;
Subsection~\ref{s:gridding}: DSP-SPARSE;
Subsection~\ref{s:adaptivity}: 
MLAS-SPARS, IMD-SPAR, SAM-SPARCS;
Subsection~\ref{s:beyondsparsity}:
MLAS-SPARS, IMD-SPAR, SAM-SPARCS;
Subsection~\ref{s:nonlinear}:
DSP-SPARSE, SAM-SPARCS;
Subsection~\ref{s:numerics}:
DSP-SPARSE;
Subsection~\ref{s:hardware}:
DSP-SPARSE, SAM-SPARCS.

\subsection{Structured sensing matrices} \label{s:structured}

Much of the theory concerning explicit performance bounds for compressive 
sensing revolves around Gaussian and other random matrices. These results
have immense value as they show us, in principle, the possibilities 
of compressive sensing. However, in reality 
we usually do not have the luxury to choose $\vA$ as we please. Instead 
the sensing matrix is often dictated by the physical properties of the 
sensing process (e.g., the laws of wave propagation) as well as by
constraints related to its practical implementability. Furthermore, 
sensing matrices with a specific structure can give rise to fast algorithms
for matrix-vector multiplication, which will significantly speed up
recovery algorithms. Thus the typical sensing matrix in practice is not 
Gaussian or Bernoulli, but one with a very specific structure,
e.g.\ see~\cite{LDS08,HS09,HBR10,PRT11}. This includes deterministic sensing
matrices as well as matrices whose entries are random variables which are
coupled across rows and columns in a peculiar way. This can make it highly
nontrivial to apply standard proof techniques from the compressive sensing 
literature.

Over the last few years researchers have developed a
fairly good understanding of how to derive compressive 
sensing theory for a variety of structured sensing matrices that arise
in applications, see for instance the survey article~\cite{DE12} for many 
examples and references as well as the work of Rauhut~\cite{Rau10}.
Despite this admirable progress, the derived bounds
obtained so far are not as strong as those for Gaussian-type random matrices.
One either needs to collect more measurements or enforce more restrictive
bounds on the signal sparsity compared to Gaussian matrices, or one has to 
sacrifice {\em universality}. Here, universality means that
a fixed (random) sensing matrix guarantees recovery of {\em all}
sufficiently sparse signals. In comparison, to obtain competitive theoretical
bounds using structured sensing matrices we may have to assume that
the locations and/or the signs of the non-zero entries of the signal
are randomly chosen~\cite{Tropp_Dictionaries,CP08,Rau10}. As a consequence 
the performance guarantees obtained are not universal, as they ``only''
hold for {\em most} signals.

So far involved and cumbersome combinatorial arguments, which need to be 
carefully adapted to the algebraic structure of the matrix for each individual 
case, often provide the best theoretical performance bounds for structured 
matrices -- and yet, as mentioned before, these bounds still fall short 
of those for Gaussian matrices. Can we overcome these limitations of the 
existing theory by developing a collection of tools that allows us to 
build a compressive sensing theory for structured matrices that is 
(almost) on par with that for random matrices?

\medskip

Now let us change our viewpoint somewhat. Assume that we {\em do} have the 
freedom to design the sensing matrix. The only condition we impose is that 
we want deterministic (explicit) constructions with the goal to establish 
performance bounds that are comparable to those of random matrices, for
instance by establishing appropriate RIP bounds.
Most bounds to date on the RIP for deterministic matrix constructions are 
based on the coherence, which in turn causes the number of required samples
to scale quadratically with the signal sparsity. In~\cite{BDF11} the
authors use extremely sophisticated and delicate arguments to achieve
an $\eps$-improvement in this scaling behavior of the bounds.

This poses the question, whether we can come up with deterministic 
matrices which satisfy the RIP in the optimal range of parameters.
It may well be that the so constructed matrices will have little
use in practice. But if we succeed in this enterprise, I expect the 
mathematical techniques developed for this purpose to have 
impact far beyond compressive sensing.

\subsection{Caught between two worlds: The gridding error} \label{s:gridding}

With a few exceptions, the development of compressive sensing until recently
has focused on signals having a sparse representation in discrete, 
finite-dimensional dictionaries. However, signals arising in applications 
such as radar, sonar, and remote sensing are typically determined by 
a few parameters in a continuous domain. 

A common approach to make the recovery problem amenable to the 
compressive sensing framework, is to discretize the continuous domain.   
This will result in what is often called the {\em gridding error} or
basis mismatch~\cite{CPS10}. By trying to mitigate the gridding error, we 
quickly find ourselves caught between two worlds, the continuous and the 
discrete world.  The issue is best illustrated with a concrete example. 
Suppose the signal of interest is a multitone signal of the form
\begin{equation}
y(t) = \sum_{k=1}^{s} c_k e^{j 2\pi t f_k},
\label{specsignal}
\end{equation}
with unknown amplitudes $\{c_k\}$ and unknown frequencies $\{f_k\}
\subset [-W,W]$.
Assume we sample $y$ at the time points $\{t_l\}_{l=1}^m \subset [0,1)$,
the goal is to find $\{f_k\}_{k=1}^s$ and $\{c_k\}_{k=1}^s$ given
$\vy:=\{y(t_1),\dots y(t_m)\}$. This is the well-known spectral estimation 
problem, and numerous methods have been proposed for its solution. But the
keep in mind that I chose~\eqref{specsignal} only for illustration purposes, 
in truth we are interested in much more general sparse signals. We choose a 
regular grid $\cG = \{\frac{\Delta i}{2W}\}_{i=-N}^{N}\subset [-W,W]$,
where $\Delta$ is the stepsize. Let the sensing matrix be 
$$\vA = [\va_1,\dots,\va_n], \qquad
\text{with $\va_i = \frac{1}{\sqrt{m}} \{e^{j2\pi t_l \Delta i/(2W)}\}_{i=-N}^N$.}$$
(An approximation to) the spectral estimation problem can now be expressed as
\begin{equation}
\vA \vx = \vy + \ve,
\label{specmodel}
\end{equation}
with $\ve = \vn +\vd$ being the error vector, where $\vn$ models additive 
measurement noise and $\vd$ represents noise due to the discretization or 
gridding error\footnote{We could also have captured the gridding error
as a perturbation $\vE$ of the sensing matrix, $\tilde{\vA}:=\vA +\vE$, but
it would not change the gist of the story.}.
Assuming that the frequencies $f_k$ fall on the grid points in $\cG$, the
vector $\vx$ will have exactly $s$ non-zero entries with
coefficients $\{c_k\}_{k=1}^s$. In general however the frequencies will not
lie on the grid $\cG$, resulting in a large gridding error, which creates a
rather unfavorable situation for sparse recovery. To guarantee
that~\eqref{specmodel} is a good approximation to the true spectral 
estimation problem, we need to ensure 
a small gridding error. For each $f_k$ to be close to some
grid point in $\cG$, we may have to choose $\Delta$ to be very small. 
However, this has two major drawbacks:
(i) the number of columns of $\vA$ will be large, which will negatively
impact the numerical efficiency of potential recovery algorithms.
(ii) The coherence of $\vA$ will be close to 1, which implies extremely bad 
theoretical bounds when applying standard coherence-based estimates.

Thus we are caught in a conundrum: Choosing a smaller discretization step
on the one hand reduces the gridding error, but on the other hand increases
the coherence as well as the computational complexity.
This problem begs for a clever solution. 

The {\em finite rate of innovation} concept~\cite{DVB07} might be useful in
this context, but that concept by itself does not lead to stable and fast 
algorithms or a framework that can handle signals that are only
approximately sparse.

Promising approaches to
mitigate the gridding problem can be found in~\cite{FW12,TBS12}.
Both of the proposed approaches have their benefits, but also some
drawbacks. Since the purpose of this paper is to point out open problems,
let me focus on the drawbacks here, but I want to stress that I find the
simplicity of~\cite{FW12} and the ingenuity of~\cite{TBS12} very appealing.
The theoretical assumptions on the signal sparsity and the dynamic range
of the coefficients in~\cite{FW12} are much more restrictive than those of
the best results we have for standard compressive sensing. Moreover, only 
approximate, but
not exact, support recovery is guaranteed. The framework of~\cite{TBS12},
based on an intriguing approach to superresolution in~\cite{CF12}, 
does not require a discretization step, but it is currently limited to 
very specific classes of signals. Also, the proposed numerical algorithm 
lacks some of the simplicity of $\ell_1$-minimization.

Can we develop a rigorous compressive sensing framework for signals which
are sparse in a continuous domain, that is applicable to a large class of
signals, and comes with simple, efficient numerical algorithms that
preserve as much as possible the simplicity and power of the 
standard compressive sensing approach? Can we derive theoretical guarantees
about the superresolution capabilities of compressive sensing based
methods? In this context we refer the reader to~\cite{AH11}, where
an infinite-dimensional framework for compressive sensing is proposed.

\subsection{Structured sparsity and other prior information} 
\label{s:adaptivity}

The work of Lustig and collaborators in MRI~\cite{LDS08} has shown that a 
careful utilization of the distribution of the large wavelet coefficients 
across scales can lead to substantial improvements in the practical 
performance of compressive sensing in MRI. 
``Classical'' compressive sensing theory does not assume any structure
or other prior information about the locations of the non-zero entries
of the signal. How can we best take advantage of the knowledge that
all sparsity patterns may not be equally likely in a signal? This question is 
a topic of active research, e.g.\ see~\cite{BCD10,EKB10,GSM12} as well
as many more references in~\cite{DE12}.

Structured sparsity is only one of many kinds of prior information
we may have about the signal or image. Besides obvious constraints
such as non-negativity of the signal coefficients, there is also 
application-specific prior information, such as the likelihood of certain 
molecule configurations or a minimum distance between sparse coefficients
due to some repelling force. In particular in the low SNR regime
the proper utilization of available prior information 
can have a big impact on the quality of the recovered signal.
The aim is to develop frameworks that can incorporate various kinds of 
prior information both at the theoretical and the algorithmic level
of compressive sensing.

\subsection{Beyond sparsity and compressive sensing} \label{s:beyondsparsity}

A very intriguing extension of compressive sensing is the problem
of recovering a low-rank matrix from incomplete information, also
known as the problem of {\em matrix completion} or {\em matrix
recovery}~\cite{Recht07,CR08}. Let $\vX$ be an $n\times n$ matrix. We
do not require that $\vX$ is a sparse matrix, but instead we assume that
most of its singular values are zero, i.e., the rank of $\vX$ is small
compared to $n$. Suppose we are given a linear map 
$\cA: \Cst^{n\times n} \to \Cst^m$ and measurements $\vy = \cA(\vX)$.
Can we recover $\vX$? Trying to find $\vX$ by minimizing the rank of $\vZ$ 
subject to $\cA(\vZ)=y$ would be natural but is computationally not
feasible. Inspired by concepts of compressive sensing we are led to consider
the {\em nuclear norm} minimization problem
$$\min \|\vZ\|_* \quad \text{subject to $\cA(\vZ) = y$},$$
where $\|\vZ\|_*$ denotes the sum of the singular values of $\vZ$.
A large body of literature has been published on the topic of matrix
completion, covering conditions and algorithms under which the nuclear
norm minimization (or variations thereof) can indeed recover $\vX$.
Interestingly, the paper~\cite{OMF11} derives a framework that allows
us to translate (some) recovery conditions from compressive sensing to 
the setting of matrix completion.

Many high-dimensional data structures are not just sparse in some basis,
but in addition are highly correlated across some coordinate axes.
For instance spectral signatures in a hyperspectral data cube are often
highly correlated across wavelength. Suppose now $\vX$ 
is a hyperspectral data matrix whose columns represent hyperspectral
images and the column index corresponds to wavelength.
We would like to acquire the information represented by $\vX$ with 
very few measurements only. We take measurements of the form
$y = \cA(\vX)$, where $\cA$ is a properly designed sensing operator.
Following ideas in~\cite{GV12} and~\cite{CS12}, it is intuitively appealing to 
combine the powers of compressive sensing and matrix completion and
consider the following optimization problem
\begin{equation}
\text{minimize} \,\, \|\vZ\|_{\ast} + \lambda \cS(\vZ) \qquad
\text{subject to}\,\, {\cal A}(\vZ) = y 
\label{CC}
\end{equation}
in order to recover $\vX$. Here the functional $\cS$ is chosen to exploit 
the sparsity inherent in $\vX$. For instance we may choose 
$$\cS(\vX) = \sum_k \|X_k\|_{TV},$$ 
where $X_k$ is the $k$-th column of $\vX$, see~\cite{GV12}. Or we could set 
$$\cS(\vX) = \|\vU \vX \vV^{\ast}\|_1,$$ 
where $\vU$ and $\vV$ are transforms designed such that $\vX$ is 
sparse with respect to the tensor basis $\vU \otimes \vV$, see~\cite{CS12}.
Clearly, many variations of the theme
are possible, cf.~\cite{WSB11,CS12} for further discussion and examples.
Optimization problems of this kind have significant potential
in a wide range of applications, such as dynamic MRI,
hyperspectral imaging, or target tracking.

All this leads us naturally to the quite ambitious task of constructing
a unifying framework that allows to make statements about the recovery 
conditions of mathematical objects that obey some minimal complexity measure 
via methods from convex optimization. An interesting step along this
lines is taken in the paper~\cite{CRP10}.
Such an undertaking must incorporate a further investigation of the 
connection between compressive sensing and information theory.
A Shannon-information theoretic analog of compressive sensing was recently
introduced by Wu and Verd\'u, see~\cite{WV10}. Further exciting results in 
this direction can be found in~\cite{DJM11,WV12}.

\subsection{Nonlinear compressive sensing} \label{s:nonlinear}

So far we have assumed that the observations we are collecting can be modeled
as {\em linear} functionals of the form 
$\langle \vx,\va_k \rangle, k=1,\dots,m$, where $\va_k^{\ast}$ is a sensing
vector representing a row of $\vA$.
However in many applications we can only take {\em nonlinear} measurements.
An important example is the case where we observe signal intensities, 
i.e., the measurements are of the form $|\langle \vx,\va_k  \rangle|^2$,
the phase information is missing. The problem is then to reconstruct $\vx$
from intensity measurements only. A classical example is the problem of
recovering a signal or image from the intensity measurements of its Fourier
transform. Problems of this kind, known as {\em phase retrieval} arise
in numerous applications, including X-ray crystallography, diffraction
imaging, astronomy, and quantum tomography~\cite{KST95}.

Concepts from compressive sensing and matrix completion have recently inspired 
a new approach to phase retrieval called {\em PhaseLift}~\cite{CSV11}.
It has been shown that if the vectors $\va_k$ are sampled
independently and uniformly at random on the unit sphere, then the
signal $\vx$ can be recovered exactly (up to a global phase factor)
from quadratic measurements by solving a trace-norm minimization problem 
provided that $m$ is on the order of $n \log n$ measurements\footnote{We
know meanwhile that in the order of $n$ measurements suffice, see~\cite{CL12}.}.
PhaseLift does not assume that the signal is sparse. 
It is natural to ask if we can extend the compressive sensing theory
to the recovery of sparse signals from intensity measurements.
Some initial results can be found in~\cite{CMB08,OYD12}, but it is clear that
this development is still in its infancy and much more remains to be done.
For instance, it would be very useful for a variety of applications
to know how many measurements are required to recover an $s$-sparse signal 
$\vx\in\Cst^n$ from Fourier-type intensity measurements.

Another type of nonlinear measurements is the case of quantized samples, 
and in the extreme case, 1-bit measurements~\cite{JLB11,PV11} 
(which is in a sense the opposite of intensity measurements).
But what about more general nonlinear measurements? For which types
of nonlinear measurements can we build an interesting and relevant 
compressive sensing theory? I expect such a potential framework to
have wide impact in disciplines like biology, where we often encounter
all kinds of nonlinear processes driven by a few parameters.

\subsection{Numerical algorithms} \label{s:numerics}

In recent years we have seen a large variety of numerical algorithms
being developed to solve various versions of the compressive sensing
problem.
While the user of compressive sensing now has a plethora of algorithms to 
choose from, a comparison of the advantages and disadvantages of individual
algorithms is difficult. Some algorithms provide guaranteed recovery
of {\em all} sufficiently sparse signals, others succeed only for many
or most signals. Some algorithms claim to be numerically efficient,
yet are only so, when very specific sensing matrices are used or
certain assumptions are fulfilled. Other algorithms are fast, but in order
to succeed they require more measurements than competing methods.
Fortunately the number of researchers who have made implementations 
of their algorithms available is large (much larger than in many
other areas where numerical algorithms play a key role), making it fairly
easy to test many of the published algorithms in a variety of scenarios.

Compressive sensing and matrix completion have stimulated the development
of a variety of efficient algorithms for $\ell_1$-minimization and 
semidefinite programming, see for instance~\cite{FNW07,HYZ08,BMM11,MGC11}.
Many of these algorithms come with rigorous theoretical guarantees.
Based on heuristic considerations, some of these algorithms have been 
extended to solve non-convex problems, 
such as $\ell_p$-minimization with $p <1$. To what extent can we support 
these promising empirical results for non-convex optimization with 
theoretical convergence guarantees?

Iterative thresholding algorithms have been proposed as numerically
efficient alternatives to convex programming for large-scale 
problems~\cite{DDM04,FR08,BT09}. 
But until recently, known thresholding algorithms have offered
substantially worse sparsity-undersampling tradeoffs than 
convex optimization. {\em Message passing algorithms} are a breakthrough in this
regard~\cite{DMM09}. Approximate Message Passing (AMP) algorithms
proposed by Donoho, Montanari and their coworkers, are low-complexity
iterative thresholding algorithms which can achieve optimal performance in
terms of sparsity-undersampling tradeoff~\cite{DJM11}.
These AMP algorithms are also able to utilize block sparsity for instance.
Interestingly, in the message passing framework of Donoho and Montanari 
we observe a shift from sparsity to the (R\'{e}nyi) information 
dimension, which in turn leads us right to the discussion at the
end of Subsection~\ref{s:beyondsparsity}.
There are also intriguing connections to statistical physics.

However, it remains a major challenge to extend the theory underlying AMP
from (Gaussian or band-diagonal) random matrices to those structured 
sensing matrices that are encountered in practice. Initial investigations
have been carried out by Schniter and collaborators, see e.g.~\cite{SS12}.

\subsection{Hardware design} \label{s:hardware}

The concept of compressive sensing has inspired the development of new 
data acquisition hardware. 
By now we have seen compressive sensing ``in action'' in a variety
of applications, such as MRI, astronomy, and
analog-to-digital conversion, see Igor Carron's list of compressive sensing
hardware~\cite{Carron}.
Yet, the construction of compressive sensing-based hardware is still a 
great challenge. 

But the process of developing
compressive sensing hardware is not the job of the domain scientist alone.
The knowledge gained during this process feeds back into the ``production
cycle'' of compressive sensing, as theoreticians (have to) learn how to
adapt their theory to more realistic scenarios, and in turn may then be able
to provide the practitioner with better insight into performance bounds 
and improved design guidelines.
Noise is a major limiting factor. Calibration remains a big problem. 
An efficient feedback loop between the different scientists 
working on theory, algorithms, and hardware design will be key to ensure
further breakthroughs in this area.

\section{The Future of Compressive Sensing}

The surest way for a scientist to make a fool of himself/herself is
by attempting to predict the future. But let me try anyway.
Is compressive sensing here to stay? How important will it be in
the future? And how will it evolve?

Where tremendous hope and a lot of enthusiasm meet, there is naturally the 
danger of a hype and thus the possibility of dramatic failure. Will the 
roadmap of compressive sensing be {\em from hope to hype to history}? 
It is clear that when we look back, say ten years from now, there will
be areas where the concept of compressive sensing was not successful.
One reason for such a failure may be that compressive sensing
seemed a promising solution as long as we looked at an isolated subproblem.
Yet, once we consider the subproblem in the context of the bigger problem
from which it was extracted, the efficiencies gained via compressive 
sensing may have diminished. 

However, I will not attempt to predict in 
which areas compressive sensing may not fulfill its promise. 
After all, if there will not be any crushed hopes, then we simply did not aim 
high enough. Or, in the words of Mario Andretti: {\em ``If everything 
seems under control, you're just not going fast enough!''}. 
Instead let me sketch some areas, where I believe that
compressive sensing will have (and in part already has had) a major impact. 

There is a growing gap between the amount of data we generate and the
amount of data we are
able to store, communicate, and process. As Richard Baraniuk points
out, in the year 2011 we produced already twice as many data as could be
stored~\cite{Bar11}. And the gap keeps widening.
As long as this development continues there is an urgent need for novel 
data acquisition concepts like compressive sensing.

There is an obvious intellectual achievement, in which compressive sensing
and sparse representations play a key role: Advanced probability theory
and (in particular) random matrix theory, convex optimization, and applied
harmonic analysis will become and already have become standard ingredients 
of the toolbox of many engineers. At the same time, 
mathematicians will have gained a much deeper understanding of how to confront
real-world applications.
Compressive sensing teaches us (or forces us?) to work across disciplines, but
not in form of an alibi collaboration whose main purpose is
to convince program directors and proposal reviewers to fund our next
``interdisciplinary'' project. No, it creates interdisciplinary
collaborations for the only sensible reason: because some important problems
simply cannot be solved otherwise! Furthermore, compressive sensing has
advanced the development of $\ell_1$-minimization algorithms, and more
generally of non-smooth optimization. These algorithms find wide-spread use
in many disciplines, including physics, biology, and economics.

There will be ``conceptual'' achievements. For example, analogously to how 
wavelet theory has taught us how to think about multiscale and sparsity 
(despite the fact that wavelets could not live up to many of our expectations),
compressive sensing will teach us how to think properly about minimal 
complexity and how to exploit it in a computationally efficient manner, and 
it may even be instrumental in developing a rigorous information theory 
framework for various areas such as molecular biology. 

To revolutionize technology we will need to develop hardware and algorithms 
via an integrated, transdisciplinary approach. 
Hence, in the future when we design sensors, processors, and other devices,
we may no longer speak only about {\em hardware} and {\em software}, where each 
of these two components is developed essentially separately. Instead, we
may have to add a third category, which we could call {\em hybridware} 
or {\em mathematical sensing}\footnote{{\em ``Mathematical Sensing , 
Dennis Healy's Vision''}, Presentation by Ronald Coifman given at the
Minisymposium ``In Memory of Dennis Healy and His Scientific Vision'', 
2010 SIAM Annual Meeting.}, where the physical device and the mathematical 
algorithm are completely intertwined and co-designed right from the beginning.

Hence, looking further into the future, maybe the most important legacy 
of compressive sensing will be that it has forced us to think about 
information, complexity, hardware, and algorithms in a truly integrated manner.

\section*{Acknowledgement}

The author would like to thank Emmanuel, Cand\`es, Michael Elad, Holger Rauhut,
and Anna Scaglione for their useful feedback on an earlier version
of this paper.

\end{document}